\documentclass[preprint]{aastex}
\usepackage{amsmath}
\usepackage{cases}
\usepackage{url}
\usepackage{color}
\usepackage{lineno}

\shorttitle{Formation of Rotational Discontinuities}
\shortauthors{Yang et al.}

\begin{document}


\title{Formation of Rotational Discontinuities in Compressive three-dimensional MHD Turbulence}


\author{Liping Yang\altaffilmark{1,2}, Lei Zhang\altaffilmark{2}, Jiansen He\altaffilmark{2},
Chuanyi Tu\altaffilmark{2}, Linghua Wang\altaffilmark{2}, Eckart
Marsch\altaffilmark{3}, Xin Wang\altaffilmark{2}, Shaohua
Zhang\altaffilmark{4}, and Xueshang Feng\altaffilmark{1}}


\altaffiltext{1}{SIGMA Weather Group, State Key Laboratory for Space
Weather, Center for Space Science and Applied Research, Chinese
Academy of Sciences, 100190, Beijing, China}
\altaffiltext{2}{School
of Earth and Space Sciences, Peking University, 100871 Beijing,
China; E-mail: jshept@gmail.com}
\altaffiltext{3}{Institute for
Experimental and Applied Physics, Christian Albrechts University at
Kiel, 24118 Kiel, Germany}
\altaffiltext{4}{Center of Spacecraft
Assembly Integration and Test, China Academy of Space Technology,
Beijing 100094, China}



\begin{abstract}
Measurements of solar wind turbulence reveal the ubiquity of
discontinuities. In this study, we investigate how the
discontinuities, especially rotational discontinuities (RDs), are
formed in magnetohydrodynamic (MHD) turbulence. In a simulation of the decaying compressive three-dimensional (3-D) MHD turbulence
with an imposed uniform background magnetic field, we detect RDs
with sharp field rotations and little variations of magnetic field
intensity as well as mass density. At the same time, in the de
Hoffman-Teller (HT) frame, the plasma velocity is nearly in
agreement with the Alfv\'{e}n speed, and is field-aligned on both
sides of the discontinuity. We take one of the identified RDs to
analyze in details its 3-D structure and temporal evolution. By
checking the magnetic field and plasma parameters, we find that the
identified RD evolves from the steepening of the Alfv\'{e}n wave
with moderate amplitude, and that steepening is caused by the nonuniformity of the Alfv\'{e}n speed in the ambient turbulence.

\end{abstract}




%
%

\section{INTRODUCTION}

Since the beginning of the space age, the solar wind has been
regarded as an excellent natural laboratory for studying the plasma
turbulence. The endeavoring research over decades has revealed that
the discontinuities are ubiquitous in the solar wind
\citep{Burlaga1969, Burlaga1971, Smith1973, Solodyna1977,
Tsurutani1979, Neugebauer1984, Lepping1986, Tsurutani1994,
Tsurutani1996, Lee1996, Horbury2001, Knetter2004, Tsurutani2005,
Neugebauer2006, Vasquez2007, Li2008, Lin2009, Sonnerup2010, Teh2011, Haaland2012,
Malaspina2012, Wang2013, Paschmann2013}. These discontinuities appear as large and
rapid changes in properties of the plasma and magnetic field, and
are identified as statically advected tangential discontinuities
(TDs), or propagating rotational discontinuities (RDs). The TDs are
characterized by small normal components of the magnetic field,
large variations of magnetic field intensity and density jumps,
while the RDs have large normal components of the magnetic field,
but small variations of magnetic field intensity and density
\citep{Hudson1970}.

By measurements of the magnetic field from Pioneer 6, \cite{Burlaga1969}
observed the discontinuities in the magnetic-field direction with
special emphasis on their distribution in time. Based on
interplanetary field measurements made by the vector helium
magnetometers onboard Pioneer 10 and Pioneer 11,
\cite{Tsurutani1979} investigated a possible dependence of the occurrence rate and the properties of the discontinuities on radial
distance between 1 and 8.5 AU.  \cite{Lepping1986} surveyed the data
from the Mariner 10 primary mission to study the characteristics of
the discontinuities in the interplanetary magnetic field at
heliographic distances of 1.0, 0.72, and 0.46 AU, and found an
$r^{-1.3\pm0.4}$ dependence for the daily average number of
discontinuities per hour. With Ulysses magnetic field and plasma
data obtained at radial distances ranging between 1 and 5 AU from the Sun and at high
heliographic latitudes, \cite{Tsurutani1994} discovered two regions
where the occurrence rate of interplanetary discontinuities is high:
in stream-stream interaction regions and in Alfv\'{e}n wave trains.
To determine the normals of the discontinuities, \cite{Horbury2001}
explored the discontinuities measured by three spacecraft WIND, IMP 8, and
Geotail together with the solar wind velocity measured at Geotail,
and obtained quite different distributions of the discontinuity
types. With magnetic field data from the ACE spacecraft,
\cite{Vasquez2007} extended the survey of discontinuity properties
to small spread angles of the field vectors across the discontinuity, and found that solar wind discontinuities
are far more abundant at small than at large spread angles. By
measurements from the WIND spacecraft, \cite{Wang2013} studied the
intermittent structures in solar wind turbulence, which are
identified as being mostly rotational discontinuities (RDs) and
rarely tangential discontinuities (TDs) based on the technique
described by \cite{Smith1973} and \cite{Tsurutani1979}. \cite{Paschmann2013} carried out a comprehensive study of directional discontinuties and Alfv\'{e}nic fluctuations in the solar wind on the basis of Cluster data.

So far, there are still debates regarding the origin and nature of the
discontinuities in the solar wind. Since RDs appear as a compressed
Alfv\'{e}n wave, nonlinear wave steepening has been suggested as the
cause of its formation \citep{Cohen1974, Malara1991, Tsurutani1994,
Vasquez1996, Vasquez1998b, Vasquez1998a, Tsurutani2002, Marsch2011}. By
numerically calculating the evolution of an initially
parallel-propagating, elliptically polarized wave train in a cold
plasma, \cite{Cohen1974} were the first to investigate this
possibility, and showed that this wave evolves to a constant-B
solution with RDs that rotate the field by exactly $180^\circ$.
 \cite{Vasquez1996} continued this study, but conducted a 1.5-D
hybrid numerical simulation study of the evolution of obliquely
propagating, linearly polarized Alfv\'{e}n wave trains. They found
that large-amplitude $dB/B_0 > 1$ wave trains steepen and produce
RDs which always rotate the field by  $<180^ \circ$.
\cite{Vasquez1998b} also presented 2.5-D numerical simulations of a
small group of nonplanar Alfv\'{e}n waves to show the generation of
imbedded RDs. It should be noted that in their models the formation of RDs probably occurs relatively near the Sun where most
Alfv\'{e}nic fluctuations originate.

There is another model suggesting that MHD turbulence
dynamically generates these discontinuities as the solar wind flows
outward. Recently, numerical simulations have been done to
investigate this assertion \citep{Greco2008, Greco2009, Greco2010,
Servidio2011, Zhdankin2012a, Zhdankin2012b}. Through analyses of MHD
simulation data, \cite{Greco2008} examined the relationship between
discontinuities identified by classical methods, and coherent
structures identified by using intermittency statistics. They found
that the two methods produce remarkably similar distributions of
waiting times, and in fact identify many of the same events.
\cite{Greco2009} further examined the link between intermittent
turbulence and MHD discontinuities, directly comparing simulations of MHD turbulence with statistical
analysis from ACE solar wind data.
Their results support the notion that some solar wind discontinuities are
consequences of intermittent turbulence. In direct numerical
simulations of MHD turbulence with an imposed uniform magnetic
field, \cite{Zhdankin2012a} investigated the statistical properties
of magnetic discontinuities, and concluded that the discontinuities
observed in the solar wind can be reproduced by MHD turbulence.
However, these works conducted statistical studies, and did not give
a clear illustration of how discontinuities, especially RDs, are formed
in MHD turbulence.

In the present study, we utilize a compressible 3-D MHD model to
illustrate and analyze the forming of RDs in the turbulence. By
checking the magnetic field and plasma properties, it is found that
RD is produced by the steepening of a moderate-amplitude Alfv\'{e}n
wave with nonuniform propagating speed. The paper is organized as
follows. In Section 2, a general description of the numerical MHD
model is given. Section 3 describes the results of the numerical
simulation and its analysis. Section 4 is reserved for the summary
and discussion.

\section{NUMERICAL MHD MODEL}

The description of the plasma is given by compressible 3-D MHD,
which involves a fluctuating flow velocity $\mathbf{v}(x, y, z, t)$,
magnetic field $\mathbf{b}(x, y, z, t)$, density $\rho(x, y, z, t)$,
and temperature $T(x, y, z, t)$. An uniform guide field $B_0$ is
assumed in the $z-$direction, so the total magnetic field is
$\mathbf{B} = \mathbf{B_0} + \mathbf{b}$.

The MHD equations are written in the following non-dimensional form:

\begin{equation}
 \frac{\partial \rho}{\partial
t}+ \nabla \cdot \rho \mathbf{u} = 0 \ ,
\end{equation}
\begin{equation}
 \frac{\partial \rho \mathbf{u}}{\partial
t}+ \nabla \cdot \left[\rho \mathbf{u} \mathbf{u} + ( p +
\frac{1}{2}\mathbf{B}^2 )\mathbf{I}-\mathbf{B} \mathbf{B}\right] = 0 \ ,
\end{equation}
\begin{equation}
 \frac{\partial e}{\partial
t}+ \nabla \cdot \left[\mathbf{u} (e + p +
\frac{1}{2}\mathbf{B}^2)-(\mathbf{u} \cdot
\mathbf{B})\mathbf{B}\right]= \nabla \cdot (\mathbf{B} \times \eta \mathbf{j}),
\end{equation}
\begin{equation}
\frac{\partial \mathbf{B}}{\partial t}+ \nabla \cdot
(\mathbf{u}\mathbf{B}-\mathbf{B}\mathbf{u}) = \eta\nabla^2
\mathbf{B} \ ,
\end{equation}
where
\begin{equation}
e=\frac 12 \rho\mathbf{u}^2+\frac{p}{\gamma-1}+\frac 12
\mathbf{B}^2, \ \ \ \ \mathbf{j}  = \nabla \times \mathbf{B}
,\end{equation} which corresponds to the total energy density and
current density, respectively. Here, $\rho$ is the mass density;
$\mathbf{u}=(v_x, v_y, v_z)$ are the $x$, $y$, and $z-$components of
velocity; $p$ is the thermal pressure; $\mathbf{B}$ denotes the
magnetic field; $t$ is time; $\gamma= 5/3$ is the adiabatic index;
and $\eta$ is the magnetic resistivity.

Three independent parameters, an initial mean density $\rho_0$, a
characteristic length $L$, and a characteristic plasma speed $v_0 =
\delta b/\sqrt{4\pi \rho_0}$ with $\delta b = \langle
b^2\rangle^{1/2}$, are used to normalize the MHD equations. Other
variables are normalized by their combinations. The dimensionless
numbers appearing in the equations are the Mach number $M_s=
v_0/c_s$, where $c_s = \sqrt{\gamma p_0/\rho_0}$ is the sound speed,
and the magnetic Reynolds number $R_m = v_0 L/\eta$. Here, we take
$M_s$ to be 0.5, consistent with the solar wind observations at 1
AU, and $R_m$ to be 1000, which is limited by the available spatial
resolution. The uniform guide field $B_0$ is two times of the
fluctuating field $\mid \delta \mathbf{b} \mid$.

We consider periodic boundary conditions in a cube with a side
length of $2\pi L$ and a resolution defined by the number of grid points which is $512^3$, and run a
simulation from an initial state with kinetic and magnetic energy
per unit mass $\langle v^2\rangle = \langle b^2\rangle =1$. The
fluctuations initially populate an annulus in the Fourier $k-$space
such that $1 \leq k \leq 8$, with constant amplitude and random
phases \citep{Matthaeus1996, Dmitruk2004}. The initial normalized
cross helicity is set to be 0.9 such that the primordial fluctuations are
highly Alfv\'{e}nic. The initial density and thermal pressure are
set to be uniform.

To solve the equations, we employ a splitting-based finite-volume
numerical scheme. The fluid part is solved by the Godunov-type
central scheme and the magnetic part by the constrained transport
approach, in conjunction with the method called second-order Monotone Upstream Schemes
for Conservation Laws (MUSCL) for reconstruction and with the
approximate Riemann solvers of Harten-Lax-van Leer (HLL) for calculation of the
numerical fluxes \citep{Feng2011}. The explicit second-order Runge-Kutta stepping with total
variation diminishing is applied in time
integration.

\section{NUMERICAL RESULTS}

\begin{figure}[htbp]
   \begin{center}
   \begin{tabular}{c}
     \includegraphics[width=6.5  in]{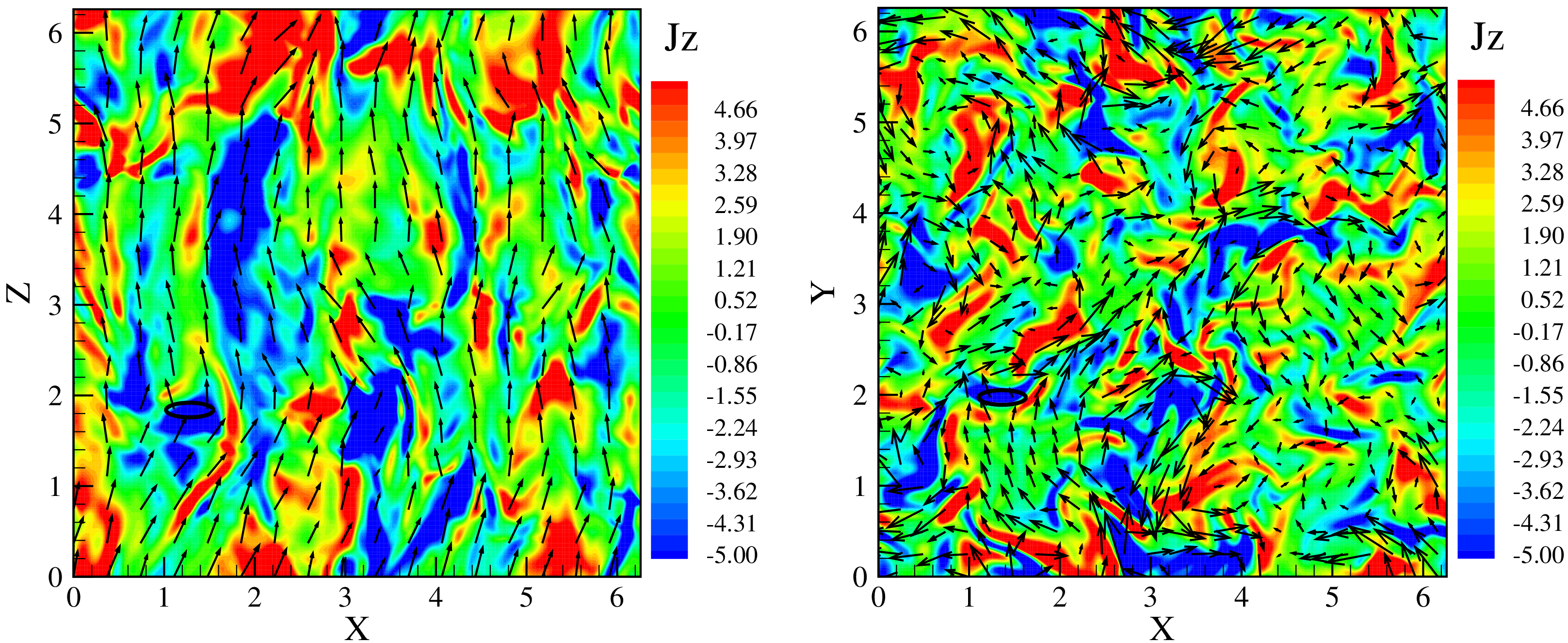}   \\
  \end{tabular}
   \end{center}
\caption{Distribution of the $z-$component of the current density $J_z$ in an $x$-$z$ (Left) and $x$-$y$ (Right) plane at
$t=1.25$. Superposed by arrows are the projections of the magnetic field vectors
in the $x$-$z$ (Left) and the $x$-$y$ plane (Right). The black ellipses mark the region where
the identified RD is formed. }\label{figure1}
\end{figure}

Figure \ref{figure1} shows the distribution of the current density in
the $z-$direction $J_z$ in the $x$-$z$ (Left) and $x$-$y$ (Right)
plane at $t=1.25$. The arrows superposed on the images are the
projections of the magnetic field vectors in the $x$-$z$ (Left) and
the $x$-$y$ plane (Right). The black ellipses mark the region where
the identified RD is formed. From this figure, we can see that a
large-scale background magnetic field is clearly present in the
$z-$direction. As a result of the well-known anisotropic behavior of
magnetic field fluctuations in MHD with an imposed uniform guide
field \citep{Matthaeus1996}, current density structures
preferentially align along the guide field direction, as shown in
the left panel, and become much more varying in the perpendicular
cross section, as shown in the right panel. Also, $J_z$ appears to
be large in magnitude at the location where the identified RD is
formed.

In order to detect RD, we first seek the regions with large
normalized partial variance of increments ($\mbox{PVI}$) of the magnetic
field vector. $\mbox{PVI}$ in 3-D space is defined as
$$\mbox{PVI}(x,y,z) = \frac{\mid \mathbf{B}(x+\Delta x, y+\Delta y, z+\Delta z) - \mathbf{B}(x-\Delta x, y-\Delta y, z-\Delta z) \mid}
{\sqrt{\langle \mid \mathbf{B}(x+\Delta x, y+\Delta y, z+\Delta z) -
\mathbf{B}(x-\Delta x, y-\Delta y, z-\Delta z) \mid ^2\rangle}}$$
where $\Delta x,\ \Delta y$ and $\Delta z$ is the grid increment in the
$x-$, $y-$ and $z-$direction, respectively. We first sample the magnetic
field $\mathbf{B}$,  plasma velocity $\mathbf{v}$, density $\rho$,
and temperature $T$ along a linear path through the region with
 $\mbox{PVI}>2$, and then perform along that path the minimum variance analysis (MVA)
\citep{Sonnerup1967} using the magnetic field data to find the
maximum variance direction ($L$), intermediate variance direction
($M$), minimum variance direction ($N$), and their corresponding
eigenvalues $\lambda_1$, $\lambda_2$, and $\lambda_3$. Finally, we
check the 3-D plasma and field structure of the possible events.

\begin{figure}[htbp]
   \begin{center}
   \begin{tabular}{c}
     \includegraphics[width=6.6 in]{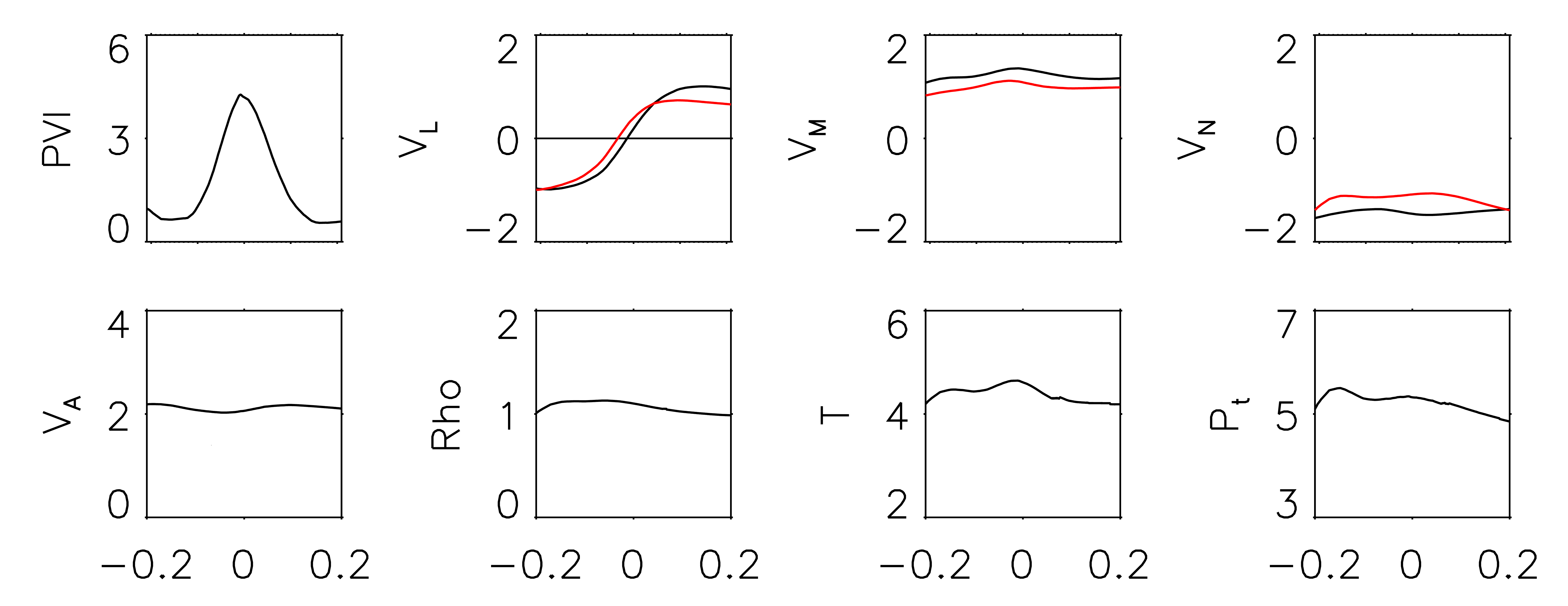}   \\
  \end{tabular}
   \end{center}
\caption{Magnetic field and plasma parameters along the sampling
interval for an identified RD. $V_L$, $V_M$, and $V_N$ denote the $L, M,
N$ components of the Alfv\'{e}n velocity (black lines) and plasma
velocity (red lines) respectively as derived from minimum variance analysis (MVA).
The Horizontal axis displays the coordinate $s$ along the sampling path, with $s=0$ at the RD point.
}\label{figure2}
\end{figure}

Figure \ref{figure2} shows the magnetic field and plasma parameters
along the sampling interval for an identified RD. In this figure,
the magnetic field has been converted into the Alfv\'{e}n velocity
$\mathbf{V_A}$, by using $\mathbf{V_A} = \mathbf{B}/ \sqrt{\rho}$ .
We can see that the sampling series of $\mbox{PVI}$ is rather close to 0,
except near the center. The large jumps of the Alfv\'{e}n velocity
and plasma velocity mainly occur along the $L$ direction, with $V_L$
jumping from -1.0 to 1.0. The $N-$components $V_N$ of them are
nearly constant, and are equal to 1.6, along with a nearly
negligible jump of $\mid \mathbf{V_A} \mid$, the magnitude of
Alfv\'{e}n speed. Also, the Alfv\'{e}n speed and plasma speed are nearly in
agreement over the entire interval, including the jump across
the RD itself. The positive correlation between them implies the RD
propagation anti-parallel to the magnetic field. The density $\rho$,
temperature $T$ and total pressure $P_t$ (which consists of the
summed magnetic pressure and thermal pressure) all exhibit
relatively constant traces throughout the interval. To be noted, the jump
of $V_L$ and the large value of $V_N$, together with the relatively
slight change of $P_t$ as well as $\rho$, corroborate that this event
is a RD.

\begin{figure}[htbp]
   \begin{center}
   \begin{tabular}{c}
     \includegraphics[width=6. in]{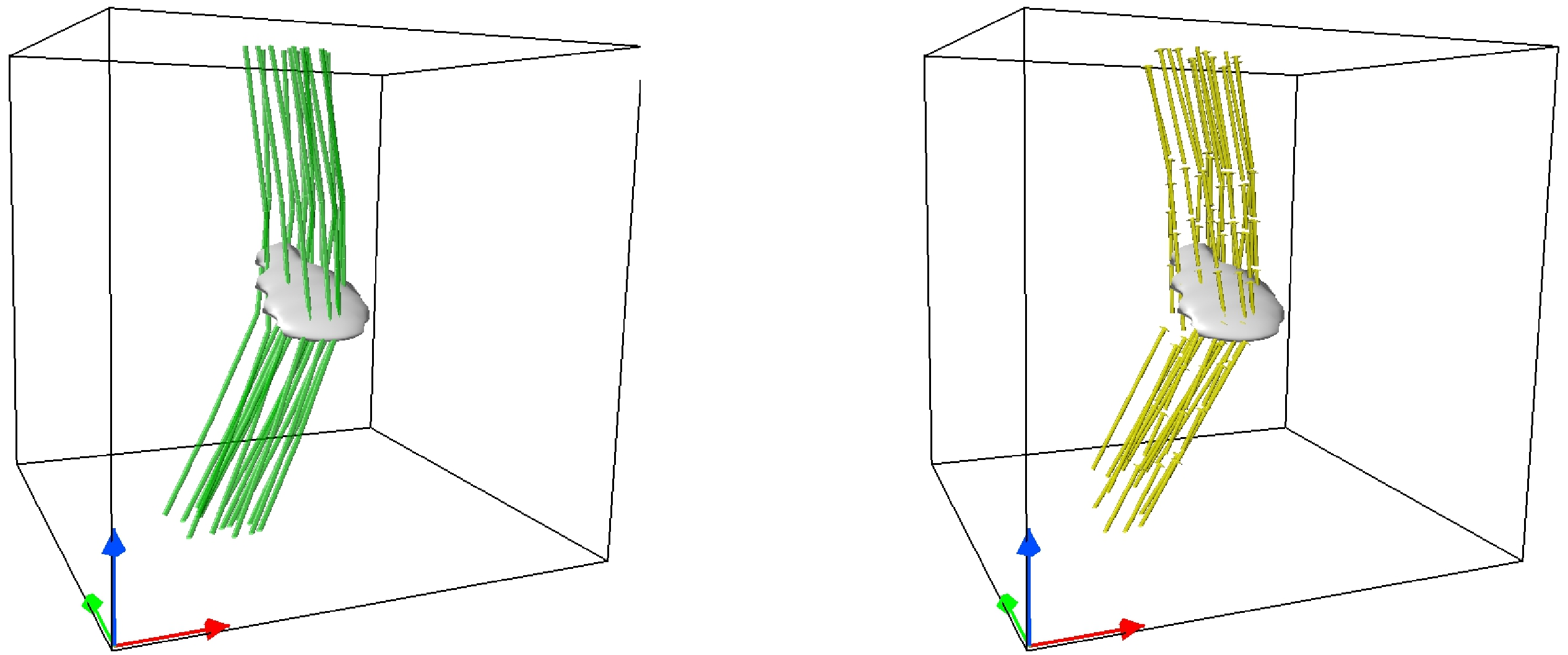}   \\
  \end{tabular}
   \end{center}
\caption{Three-dimensional structure of the identified RD. The green
lines in the left panel denote magnetic field lines and the yellow arrows
in the right panel are plasma velocity vectors, which are converted into
de Hoffman-Teller (HT) frame. The light gray surface is the
isosurface where $\mbox{PVI} = 4$, showing the RD, and red, green as well as blue
arrows display the $x-$, $y-$, and $z-$direction, respectively.}
\label{figure3}
\end{figure}

Figure \ref{figure3} exhibits the 3-D structure of the identified
RD. The green lines in the left panel denote magnetic field lines and the
yellow arrows in the right panel are plasma velocity vectors, which are
converted into the de Hoffman-Teller (HT) frame. The light gray
surface is the isosurface where $\mbox{PVI} = 4$, showing RD, and red, green as
well as blue arrows display the $x-$, $y-$, and $z-$direction,
respectively. This figure displays that in the HT frame, magnetic field
and plasma velocity have evident normal components across the RD,
and they both rotate by a certain angle. Also, the plasma velocity
is field aligned on both sides of the discontinuity, in accordance
with the Alfv\'{e}nic nature of RD. The identified RD appears as a
thin surface and its normal inclines to the $z-$axis.

\begin{figure}[htbp]
   \begin{center}
    \begin{tabular}{c}
     \includegraphics[width=6.5 in]{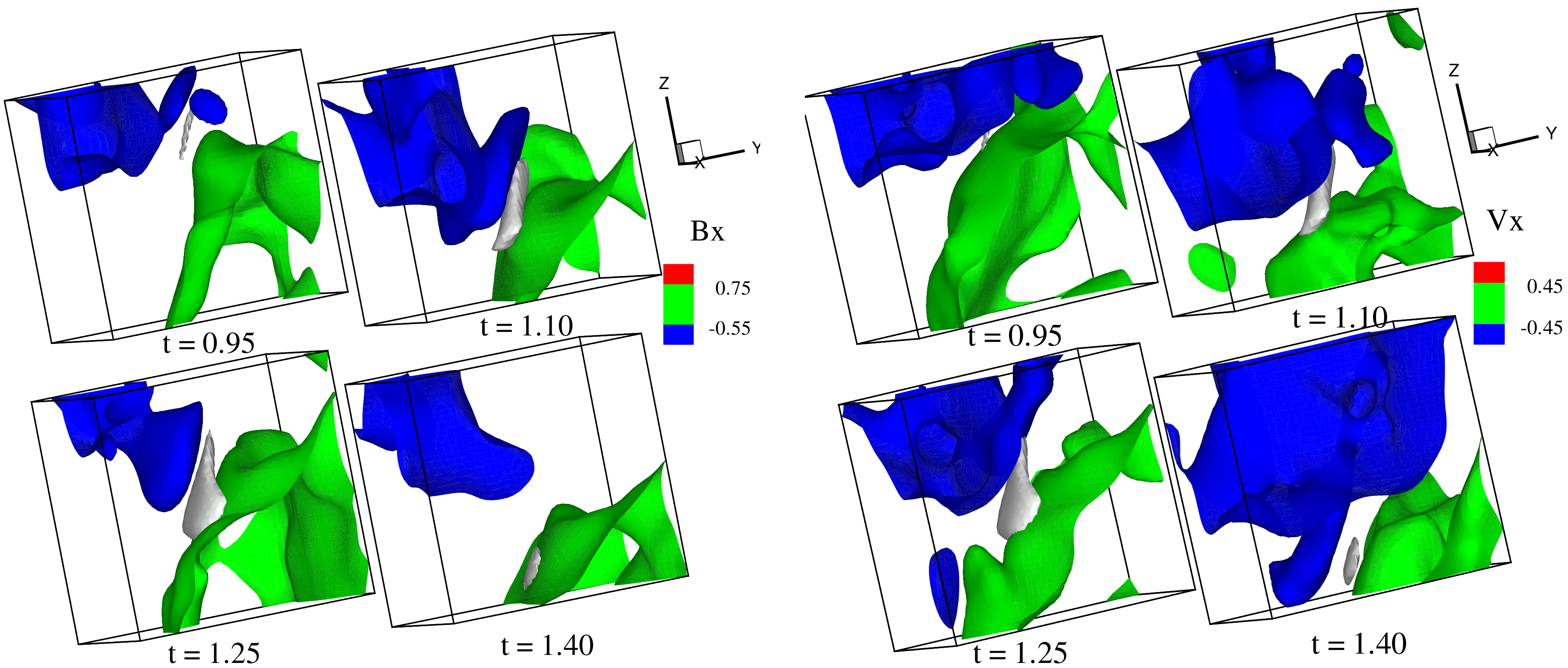}   \\
  \end{tabular}
   \end{center}
\caption{The isosurfaces of $B_x$ (Left) and $V_x$ (Right) at
different moments in time, with the light gray surface being the
isosurface where $\mbox{PVI} = 3$, and exhibiting the identified
RD.}\label{figure4}
\end{figure}

To see how the identified RD is formed, Figure \ref{figure4} shows
the isosurfaces of $B_x$ (Left) and $V_x$ (Right) at different moments in time. These isosurfaces are associated with the
Alfv\'{e}n wave as shown below. $\mbox{PVI} = 3$ is used here to exhibit
the identified RD at these four moments, and is shown as the
light gray surface. From this figure, it is notable that the mutual
approaching of the two isosurfaces of $B_x$ ($V_x$), which describes the
steepening of the Alfv\'{e}n wave, leads to the formation of a RD. At $t
= 0.95$, the isosurface with $B_x = -0.55$ ($V_x = -0.45$) is relatively
away from the isosurface with $B_x = 0.75$ ($V_x = 0.45$), and the
transition of $B_x$ ($V_x$) from $B_x = -0.55$ ($V_x = -0.45$) to
$B_x = 0.75$ ($V_x = 0.45$) is gentle. The isosurface where $\mbox{PVI} = 3$ is
small and thin. At $t = 1.10$, the isosurfaces with $B_x = -0.55$ ($V_x =
-0.45$) and $B_x = 0.75$ ($V_x = 0.45$) approach each other, and the
transition between the two isosurfaces of $B_x$ ($V_x$) becomes
steep. Correspondingly, the isosurface where $\mbox{PVI} = 3$ grows. Afterwards,
that is at $t = 1.25$, the two isosurfaces of $B_x$ ($V_x$) are
close enough to each other, and the transition between the two
isosurfaces of $B_x$ ($V_x$) becomes steeper. The isosurface where $\mbox{PVI} =
3$ grows larger and thicker, and the identified RD is fully grown.
However, at $t = 1.40$, the isosurfaces with $B_x = -0.55$ ($V_x =
-0.45$) is far away from the isosurface with $B_x = 0.75$ ($V_x = 0.45$),
and the transition between the two isosurfaces of $B_x$ ($V_x$)
becomes gentle again. As a result, the isosurface where $\mbox{PVI} = 3$ becomes
small and thin. The identified RD starts to collapse.

\begin{figure}[htbp]
   \begin{center}
    \begin{tabular}{c}
     \includegraphics[width=6. in]{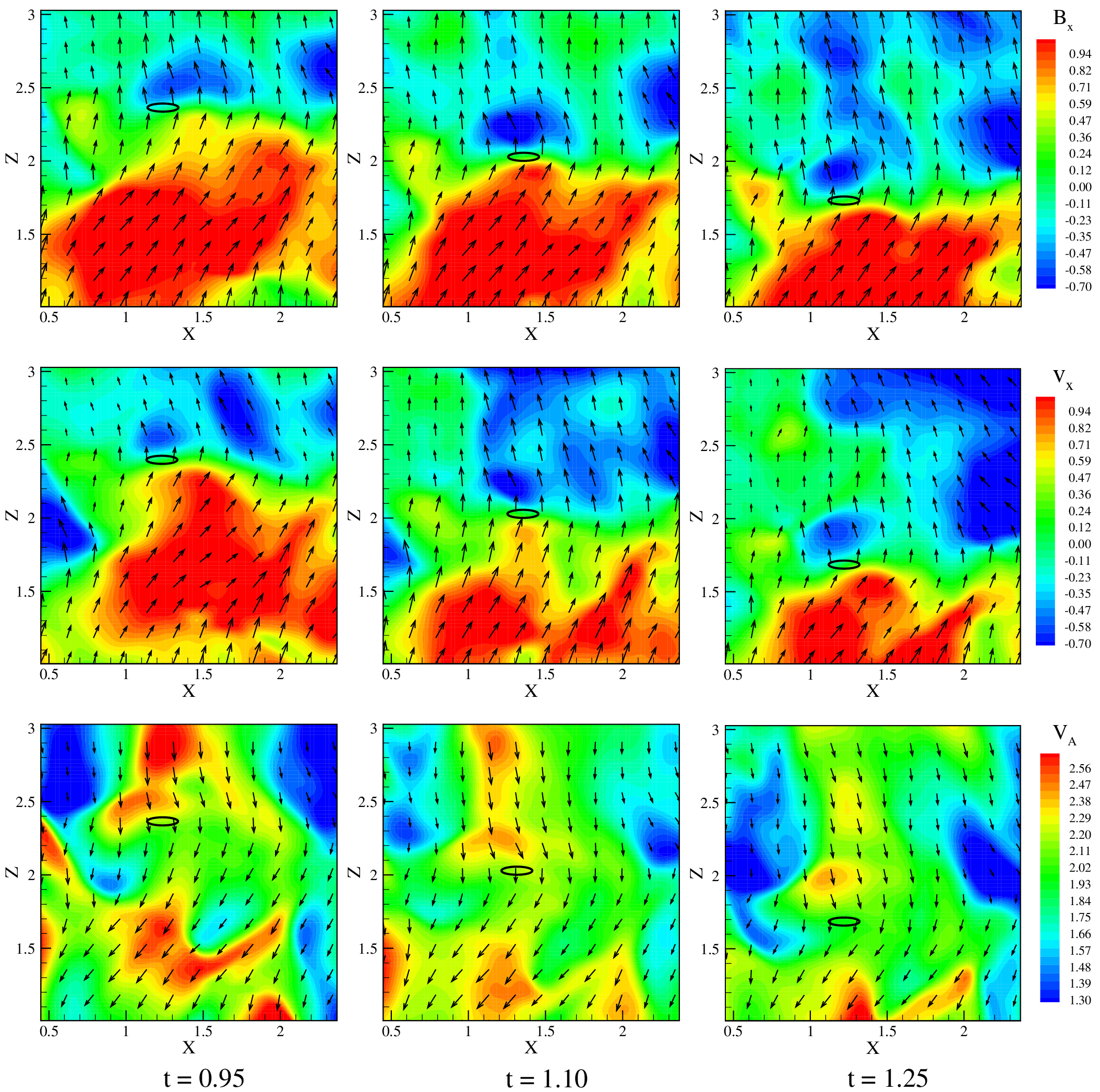}   \\
  \end{tabular}
   \end{center}
\caption{Distribution of $B_x$ (the first row), $V_x$ (the second
row), and $V_A$ (the third row) in a subzone of the $x$-$z$ plane
(which passes through the identified RD) at different moments in time.
Superposed by arrows are the projections of the magnetic field
(the first row), velocity field in the HT frame (the second row), and
negative Alfv\'{e}n speed field (the third row) vectors in the
$x$-$z$ plane. The black ellipses mark the position where the
identified RD is formed. }\label{figure5}
\end{figure}

To understand the type of waves before the RD is formed and to
investigate the process of the evolution of the isosurfaces
mentioned above, Figure \ref{figure5} exhibits the distribution of $B_x$
(the first row), $V_x$ (the second row), and $V_A$ (the third row)
in the neighborhood of the identified RD in the plane $x$-$z$ at
different moments in time. Superposed by arrows are the projections of
the magnetic field (the first row), velocity field in the HT frame
(the second row), and negative Alfv\'{e}n speed field (the third
row) vectors in the $x$-$z$ plane. The black ellipses mark the
position where the identified RD is formed. Near the ellipses, the
magnitudes of $B_x$ and $V_x$ are almost same (they are set to have
the same color scales so that the same color stands for the same
value), and the directions of the in-plane projection of the $\mathbf B$
and $\mathbf V$ vectors are almost identical. Hence in this
vicinity, we have $\mathbf V = \mathbf B / \sqrt{\rho_0}$ (we recall that
$\rho_0 \approx 1$), which agrees with the polarity relations of an
Alfv\'en wave. In other words, the RD is detected in a neighbouring Alfv\'enic environment that apparently favours RD formation.

Therefore, $V_A$ can be regarded as the propagation velocity of the
structures relative to the location where the RD forms. Hence it is
significant to trace the changes of the Alfv\'{e}n speed. In the third
row in Figure \ref{figure5} where the evolution of the Alfv\'{e}n speed is shown, we can see
that at $t = 0.95$, there is a difference of Alfv\'{e}n speed across
the black ellipse, which makes the layers with negative $B_x$ (blue
in Figure \ref{figure4}), propagate faster than that with positive
$B_x$ (green in Figure \ref{figure4}). At $t = 1.10$, there is the
evidence for approaching and squeezing of these two layers. The
difference of Alfv\'{e}n speed there remains. This
will drive these two layers further to approach each other. At $t =
1.25$, their transition becomes sharp, and the
difference of Alfv\'{e}n speed nearly
disappears. This status continues until $t = 1.40$, when the
transition becomes gentle as a result of the
faster propagation of the layers with positive $B_x$ than that
with negative $B_x$. It is obvious that the difference of
Alfv\'{e}n speed makes the Alfv\'en wave steepen, a process which forms the identified RD.

\section{SUMMARY AND DISCUSSION}

In this study, we use a simulation of the decaying compressive 3-D MHD
turbulence with an imposed uniform guide field as a test case to
explore the formation of RDs in MHD turbulence. Motivated by solar wind
observation at 1 AU, we consider a moderate fluctuation amplitude corresponding to
$\delta b / B_0 = 0.5$ and high Alfv\'{e}nic correlations with the
normalized cross helicity initially of 0.9. A case study is thus
conducted to illustrate the origin of RDs in MHD turbulence.

The numerical simulation shows the well-known anisotropic behavior
of the turbulent MHD field, with the current density structures
preferentially aligning along the guide field direction and
scattering in the perpendicular plane. To detect the RDs in this
simulated  magnetofluid, we first seek the regions with large $\mbox{PVI}$, then
conduct a MVA by sampling the parameters along a linear path, and
finally check the 3-D structure of the possible RD events. One clear RD is
identified with sharp field rotations and little variations of the
magnetic field intensity as well as density. At the same time, in
the HT frame, the plasma velocity is nearly in agreement with the
Alfv\'{e}n speed, and is field aligned on both sides of the
discontinuity, satisfying the Walen relation that expresses the Alfv\'{e}nic nature of an RD. The normal
direction of the identified RD inclines to the $z-$axis, and
propagates anti-parallel to the guide field.

The comprehensive information obtained by the simulation of the magnetic field and plasma parameters
associated with the RD implies that the RD is produced by the steepening
of the moderate-amplitude Alfv\'{e}n wave with nonuniform Alfv\'{e}n
speed in the ambient turbulence. Before the RD is formed, the layers with
negative $B_x$ smoothly transits to the layers with positive $B_x$. However, there
is a difference of Alfv\'{e}n speed across them, which
makes the layers with negative $B_x$ chase after its counterpart with
positive $B_x$. As they are driven by the neigbouring turbulence to approach and squeeze each
other, the transition between them undergoes further
steepening until the difference of Alfv\'{e}n speed nearly
disappears. At the same time, the identified RD is formed.

In this work we investigated only one RD. Certainly, there
are many RDs generated in the simulation of decaying MHD turbulence.
It will be worthwhile to conduct statistical studies to see
whether all RDs are produced by this nonlinear steepening of an
Alfv\'{e}n wave. Meanwhile, TDs are also formed. It can be inspiring to investigate their
generation mechanism in MHD turbulence. Furthermore,
the parameters we adopted in the simulation (e.g.~Mach number,
cross helicity, plasma $\beta$) may influence the forming of
RDs or TDs. In the future, we plan to conduct a parameter study to
investigate the possible effects induced by parameter variation on the generation of discontinuties in MHD turbulence.



\begin{acknowledgments}

This work is supported by NSFC under contract
Nos. 41304133, 41474147, 41231069, 41222032, 41174148, 41421003, and
41204105. The numerical calculation has been completed on computing
system at Peking University.

\end{acknowledgments}

\bibliographystyle{apj}
\bibliography{all}

\end{document}